\documentclass[preprint]{aastex631}

\usepackage{amsmath}
\usepackage{graphicx}
\usepackage{multirow}
\usepackage[multiple]{footmisc}
\usepackage{hyperref}
\usepackage[nameinlink]{cleveref}
\usepackage{xcolor}
\newcommand{\rev}[1]{#1}

\crefname{paragraph}{\S}{\S\S}
\shorttitle{OGLE-2014-BLG-0676} 
\shortauthors{Idei et al.}

\begin{document}

\title{Characterizing Microlensing Planetary System OGLE-2014-BLG-0676L with High-Resolution Image Constrained Light Curve Modeling}

\author[0009-0005-7072-1334]{Asahi Idei}
\affiliation{Department of Earth and Space Science, Graduate School of Science, The University of Osaka, Toyonaka, Osaka 560-0043, Japan}

\author[0000-0003-2302-9562]{Naoki Koshimoto}
\affiliation{Department of Earth and Space Science, Graduate School of Science, The University of Osaka, Toyonaka, Osaka 560-0043, Japan}

\author[0000-0002-5843-9433]{Daisuke Suzuki}
\affiliation{Department of Earth and Space Science, Graduate School of Science, The University of Osaka, Toyonaka, Osaka 560-0043, Japan}

\author[0009-0005-3414-455X]{Kansuke Nunota}
\affiliation{Department of Earth and Space Science, Graduate School of Science, The University of Osaka, Toyonaka, Osaka 560-0043, Japan}

\author[0000-0001-8043-8413]{David P. Bennett}
\affiliation{Laboratory for Exoplanets and Stellar Astrophysics, NASA/Goddard Space Flight Center, Greenbelt, MD 20771, USA}
\affiliation{Department of Astronomy, University of Maryland, College Park, MD 20742, USA}
\affiliation{Center for Research and Exploration in Space Science and Technology, NASA/GSFC, Greenbelt, MD 20771, USA}

\author[0000-0002-8131-8891]{Ian A. Bond}
\affiliation{Institute of Natural and Mathematical Sciences, Massey University, Auckland 0745, New Zealand}

\author[0000-0003-0014-3354]{Jean-Philippe Beaulieu}
\affiliation{School of Natural Sciences, University of Tasmania, Private Bag 37 Hobart, Tasmania, 7001, Australia}
\affiliation{Sorbonne Universit\'e, CNRS, Institut d’Astrophysique de Paris, IAP, F-75014, Paris, France}

\author[0000-0002-4035-5012]{Takahiro Sumi}
\affiliation{Department of Earth and Space Science, Graduate School of Science, The University of Osaka, Toyonaka, Osaka 560-0043, Japan}

\author[0000-0002-4746-2953]{Aparna Bhattacharya}
\affiliation{Laboratory for Exoplanets and Stellar Astrophysics, NASA/Goddard Space Flight Center, Greenbelt, MD 20771, USA}
\affiliation{Department of Astronomy, University of Maryland, College Park, MD 20742, USA}
\affiliation{Center for Research and Exploration in Space Science and Technology, NASA/GSFC, Greenbelt, MD 20771, USA}

\author[0000-0001-5860-1157]{Joshua W. Blackman}
\affiliation{Physikalisches Institut, Universit\"at Bern, Gessellschaftsstrasse 6, CH-3012 Bern, Switzerland}

\author[0009-0007-1381-3384]{Ryusei Hamada}
\affiliation{Department of Earth and Space Science, Graduate School of Science, The University of Osaka, Toyonaka, Osaka 560-0043, Japan}

\author[0009-0001-9104-0328]{Tsutsumi Nagai}
\affiliation{Astronomical Science Program, The Graduate University for Advanced Studies, SOKENDAI, 2-21-1 Osawa, Mitaka, Tokyo 181-8588, Japan}

\author[0009-0007-7197-483X]{Takuto Tamaoki}
\affiliation{Department of Earth and Space Science, Graduate School of Science, The University of Osaka, Toyonaka, Osaka 560-0043, Japan}

\author[0000-0002-5029-3257]{Sean K. Terry}
\affiliation{Laboratory for Exoplanets and Stellar Astrophysics, NASA/Goddard Space Flight Center, Greenbelt, MD 20771, USA}
\affiliation{Department of Astronomy, University of Maryland, College Park, MD 20742, USA}
\affiliation{Center for Research and Exploration in Space Science and Technology, NASA/GSFC, Greenbelt, MD 20771, USA}

\author[0000-0002-9881-4760]{Aikaterini Vandorou}
\affiliation{Laboratory for Exoplanets and Stellar Astrophysics, NASA/Goddard Space Flight Center, Greenbelt, MD 20771, USA}
\affiliation{Department of Astronomy, University of Maryland, College Park, MD 20742, USA}
\affiliation{Center for Research and Exploration in Space Science and Technology, NASA/GSFC, Greenbelt, MD 20771, USA}


\begin{abstract}

We present an analysis that incorporates high-resolution Keck adaptive optics (AO) imaging into microlensing light-curve modeling for the planetary microlensing event OGLE-2014-BLG-0676.
Using Keck AO observations obtained 6.3 years after the event, we directly resolved the lens and source.
The Keck images reveal a tension, in that the $K$-band source flux is $0.52 \pm 0.22$ magnitudes brighter than predicted by previously reported light-curve models.
By incorporating the Keck imaging constraints into the light-curve modeling, 
\rev{
we find a host star mass of $M_{\rm host} = 0.60^{+0.17}_{-0.14}\,M_{\odot}$, 
a lens distance of $D_{\rm L} = 1.88^{+0.63}_{-0.35}$~kpc, 
a planet mass of $m_{\rm p} = 3.11^{+1.11}_{-0.63}\,M_{\rm J}$, 
and a projected separation of $a_{\perp} = 2.04^{+0.44}_{-0.35}$~au and $a_{\perp} = 3.72^{+0.92}_{-0.72}$ au for the close and wide solution, respectively.
}
These results demonstrate the power of combining high-angular-resolution imaging with microlensing light-curve modeling to mitigate potential systematic effects and modeling degeneracies, enabling robust determinations of the physical properties of microlensing planetary systems.
The results presented here can be confirmed by future observations from the \textit{Roman}'s Galactic Plane Survey.

\end{abstract}



\section{Introduction}
\label{sec:intro}

The gravitational microlensing method \citep{mao91} is currently the only exoplanet detection method that is sensitive down to Earth-mass planets \citep{ben96} located beyond the snow line \citep{gouloe92}. 
To date, this method has led to the discovery of over 200 exoplanets \citep{ps, ake13}.
For these planets, statistical analysis has revealed a variety of interesting results, for example, the exoplanet mass-ratio function \citep{suzu16, cas12, sum10, zan25} and the dependence of microlensing planet frequency on the host star mass \citep{koshimoto2021, nuno24}.

On the other hand, there are some observational limitations. 
One of the major limitations of microlensing is that the only universally measurable quantity in all events is the Einstein radius crossing time, $t_{\rm E}$. 
This timescale is defined as
\begin{equation}
    t_{\rm E} = \frac{\theta_{\rm E}}{\mu_{\rm rel}} = \frac{1}{\mu_{\rm rel}}\sqrt{\kappa M_{\rm L} \left(\frac{1}{D_{\rm L}} - \frac{1}{D_{\rm S}}\right) \rm AU}
\end{equation}
where $\theta_{\rm E}$ is the Einstein angular radius, $\mu_{\rm rel}$ is the lens–source relative proper motion, $\kappa = 8.144 ~{\rm mas}~M^{-1}_{\odot}$, $M_{\rm L}$ is the lens mass, $D_{\rm L}$ is the distance from the observer to the lens, and $D_{\rm S}$ is the distance from the observer to the source. 
As this equation shows, $t_{\rm E}$ alone is insufficient to uniquely determine the physical properties of the lens, such as $M_{\rm L}$ and $D_{\rm L}$. 
To break this degeneracy, at least two additional higher-order effects, such as the microlensing parallax \citep{gou92} and finite source effect \citep{witmao94} are required.
However, they share two significant limitations: first, higher-order effects do not always provide accurate or unique constraints on the lens mass and distance; and second, because microlensing events are transient, such effects can only be measured during the event and cannot be re-measured afterward.

Therefore, in events where higher-order effects cannot be measured, or in cases where light-curve models remain highly degenerate despite the detection of such effects (e.g., MOA-2010-BLG-328; \citealt{fur13}), an alternative approach is required to determine the physical properties of the lens system.
In such cases, a common approach is to perform a Bayesian analysis, in which prior distributions of lens and source properties, based on Galactic models, are combined with the light-curve information to statistically estimate the lens mass and distance \citep[e.g.,][]{ben14}. 
This method provides probabilistic constraints even when no higher-order microlensing effects are measured.
An alternative method was theoretically proposed by \citet{ben07}, who showed that high angular resolution imaging obtained several years after the microlensing event can spatially resolve the lens and source stars, unless the lens is a stellar remnant \citep{bla21}.
When combined with the microlensing timescale measured from the light curve, such imaging provides additional constraints that enable the determination of the lens mass and distance without relying solely on higher-order microlensing effects.
Since the light-curve modeling yields the planet-to-host mass ratio $q$, the planet mass can then be derived by combining $q$ with the independently constrained lens mass.

Several measurements of the lens system masses have been made using this technique.
One prominent example is OGLE-2005-BLG-169, for which high-angular-resolution follow-up observations successfully resolved the source and lens stars \citep{bat15, ben15}.
This event was observed with both the \textit{Hubble Space Telescope} and the NIRC2 adaptive optics (AO) system on the Keck telescope.
These observations demonstrated the effectiveness of the method, enabling direct measurements of the masses and distance of the planetary system.
The result revealed a Uranus-mass planet orbiting a main-sequence host star located at a distance of approximately 4~kpc.
Many other successful mass measurements based on high-angular-resolution follow-up imaging have now been published (e.g., \citealt{bha18, ben20, van20, ter24}).

Here we present another example of a planetary microlensing event, OGLE-2014-BLG-0676/MOA-2014-BLG-175 (hereafter OB140676).
This event is located at (R.A., decl.) (J2000) = (17$^{\mathrm h}$52$^{\mathrm m}$24$^{\mathrm s}$.50, $-$30$^{\circ}$32$^{\mathrm m}$54$^{\mathrm s}$.20), where the lens–source angular separation was able to be resolved by the follow-up high angular resolution imaging.
OB140676 was discovered and monitored by the microlensing survey telescopes of the Optical Gravitational Lensing Experiment (OGLE) \citep{uda15} and the Microlensing Observations in Astrophysics (MOA) collaborations \citep{bon01, sum03}. 
This planetary event was previously analyzed in detail and the microlensing parameters were reported by \citet{rat17}. 
Their light curve modeling detected the finite source effect, allowing them to measure the angular Einstein radius as $\theta_{\rm E} = 1.38 \pm 0.43$ mas. 
However, no significant microlensing parallax signal was found in their analysis. 
They estimated the lens mass and distance using a Bayesian analysis based on a Galactic model, obtaining $M_{\rm L} = 0.62^{+0.20}_{-0.22}\,M_{\odot}$ and $D_{\rm L} = 2.22^{+0.96}_{-0.83}\,\mathrm{kpc}$. 
From the planet-to-host mass ratio derived in the light curve fit, they inferred a planet mass of $M_{\rm p} = 3.09^{+1.02}_{-1.12}\,M_{\rm J}$. 
Although the lens mass and distance were not directly derived from the light curve modeling, the $\mu_{\rm rel} = 4.3 \pm 1.4\,{\rm mas/yr}$ implies that the lens and source could be resolved within 10 years.

\citet{xie21} conducted high-resolution imaging of the event approximately 1--2~years after its peak using the Keck and Magellan telescopes. 
At that epoch, the angular separation between the lens and the source was too small to be resolved. 
Therefore, they measured the total flux of the target in the $K$ band (Magellan) and $J$ band (Keck), and then subtracted the source flux estimated from the static light-curve model of \citet{rat17}.
The remaining blended flux was assumed to originate entirely from the lens star, from which they inferred the lens mass and distance.
Although this approach provided the observational constraints from high-resolution imaging, it left several important uncertainties unresolved.
Because the lens and source were not yet separated, the inferred parameters remained highly uncertain, and the assumption that blended light was lens-originated may not necessarily hold true. 
Also, the source flux they used depends on the light curve modeling. 
To overcome these issues, additional observations taken at a later epoch---when the lens and source are sufficiently separated---are essential. 
Such observations would enable direct photometric measurements of both the lens and the source, thereby allowing the lens properties to be determined without relying on the assumptions adopted in \citet{xie21}.

In this paper, we present a new high-resolution imaging analysis of the microlensing event OB140676, using Keck-I/OSIRIS $K_p$-band observations obtained about 6.3~years after the event.
At this epoch, the lens and source were clearly resolved, enabling direct measurements of their individual fluxes and relative proper motion.
This paper is organized as follows. 
In Section \ref{sec:Keck}, we describe the Keck high angular resolution follow-up observations and their analysis, and in Section \ref{sec:murel}, we discuss inconsistencies with the previous light curve analysis. 
Section \ref{sec:relight} shows how we conduct the light curve modeling with Keck image constraints.
In Section \ref{sec:result}, we present the lens system mass and distance that are measured by the combined light curve and Keck follow-up data. 
Finally, we discuss the implications of these results and present our conclusions in Section \ref{sec:discus} and Section \ref{sec:conclusion}.


\section{Keck Follow up Observations and Analysis}
\label{sec:Keck}

\subsection{Observations and Reduction}
\label{sec:keck_data}  
We observed OB140676 using the Keck-I/OSIRIS camera in the ${K_p}$-band with the laser guide star adaptive optics system on August 10, 2020 (MJD = 59071.32), which is 2294.48 days (6.28 years) after the time of maximum magnification. 
\rev{
The measured pixel scale of the OSIRIS camera is 9.952 mas pixel$^{-1}$ 
\citep[e.g.,][]{fre23}, with each image consisting of $2048 \times 2048$ pixels. 
Note that we used the designed pixel scale of 10 mas pixel $^{-1}$ in the analysis, but the difference does not affect our result as it is much smaller than the uncertainties of the measurements.
}
A total of ten science images were obtained by dithering exposures, each with an integration time of 59.011 seconds. 
\rev{
A nearby natural guide star with $R = 16.24$ mag, located $9.82\arcsec$ from the target, was used for tip-tilt correction.
}
In addition, dark, flat, and sky images were acquired. 
Since no dark and flat images were available on the same day as the science image, those taken on adjacent days were selected. 
For the sky images, we chose those taken on the same day as the science image and the previous day.

The data reduction was performed using the Keck AO Imaging (KAI) data reduction pipeline \citep{lu22}.
The pipeline includes image registration, flat-field correction, dark subtraction, and masking of bad pixels and cosmic rays, resulting in a final combined image for analysis. 
Of the ten original images, four in which the stellar point spread function (PSF) FWHM exceeded 70 mas were excluded to ensure data quality.
The remaining six images were combined to produce the co-added image shown in the top-left panel of Figure~\ref{fig:quad-panel}.
\rev{
The image quality of the selected frames is characterized by FWHM values ranging from 58.8 to 68.3 mas, with a median value of 64 mas. 
The corresponding Strehl ratios range from 0.19 to 0.30, with a median of 0.22.
}

\rev{
The target is shown in the top-right panel of Figure~\ref{fig:quad-panel}. 
For a clearer view of the PSF structure, we also provide a comparison between the target and a nearby isolated single star in Appendix~\ref{app:single_vs_traget}.
}

\subsection{PSF Photometry}
\label{sec:keck_phot}
We identified the target by comparing our image with the previous high-angular-resolution images observed by \citet{xie21}.
We found that the target PSF is slightly elongated compared to those of nearby isolated stars, suggesting that the source and lens stars are partially resolved.
In such a case, the point spread function (PSF) photometry is necessary to measure the position and brightness of the individual stars precisely.
We use the \texttt{DAOPHOT} software \citep{ste87} to make an empirical PSF model and then use the \texttt{emcee} ensemble sampler \citep{for13} to perform the Markov Chain Monte Carlo (MCMC) method for computation of the posterior distribution of the source and lens positions and brightness using the PSF model.

The PSF model construction process consists of three stages. First, we used the \texttt{FIND} and \texttt{PHOTOMETRY} (aperture photometry) commands in \texttt{DAOPHOT} to identify all potential stars in the image and estimate the sky background for each star. 
Next, we employed the \texttt{PICK} command to select candidate stars for the construction of the PSF model.
We applied additional criteria to determine the final set of PSF stars, 
selecting stars with instrumental magnitudes $K_{\rm star}<14.5$ within a 500-pixel radius of the target,
excluding those with a neighboring star brighter than $K_{\rm star}+3\ \rm mag$ within the PSF RADIUS (15 pixels),
to avoid contamination in the PSF model construction.
Using these criteria, we selected 16 stars.
We then visually inspected the images and excluded stars whose PSF profiles were clearly different from those of other stars or were affected by blending, resulting in a final sample of 14 stars.

The last step for the PSF construction is to perform the \texttt{PSF} command using the selected PSF stars.
\texttt{DAOPHOT} constructs the PSF model as a hybrid of an analytic PSF and an empirical correction. 
The analytic PSF is first fitted to the PSF stars using pixel values within a specified radius, \texttt{FITTING RADIUS}. 
The look-up table stores the residuals from this fit beyond \texttt{FITTING RADIUS}, up to \texttt{PSF RADIUS}, allowing the empirical component to capture deviations that the analytic model alone cannot fully describe.
To determine the optimal PSF model, we explored different parameter settings. 
The functional form of the analytic PSF model was selected automatically by setting \texttt{ANALYTIC MODEL PSF} to -7, allowing \texttt{DAOPHOT} to test multiple models—including Gaussian, Moffat, Lorentz, and Penny functions—and choose the one that minimized the residual scatter (``chi") within the \texttt{FITTING RADIUS}. 
Additionally, we examined the effect of \texttt{VARIABLE PSF}, which controls whether the empirical PSF correction varies across the image. 
Setting this parameter to 0 assumes a spatially invariant PSF, while values of 1 and 2 allow for a linear or quadratic variation with position, respectively.
The evaluation of the PSF model was based on the residuals from the final hybrid PSF model, which accounts for both the analytic and empirical components. 
This is to assess the overall effectiveness of the combined model and is different from the \texttt{DAOPHOT} output, ``chi", which assesses just the analytic component. 
After testing different configurations, we adopted \texttt{FITTING RADIUS} = 5 pixels and \texttt{VARIABLE PSF} = 2. 

\begin{figure*}
    \centering
    \includegraphics[width=0.6\linewidth]{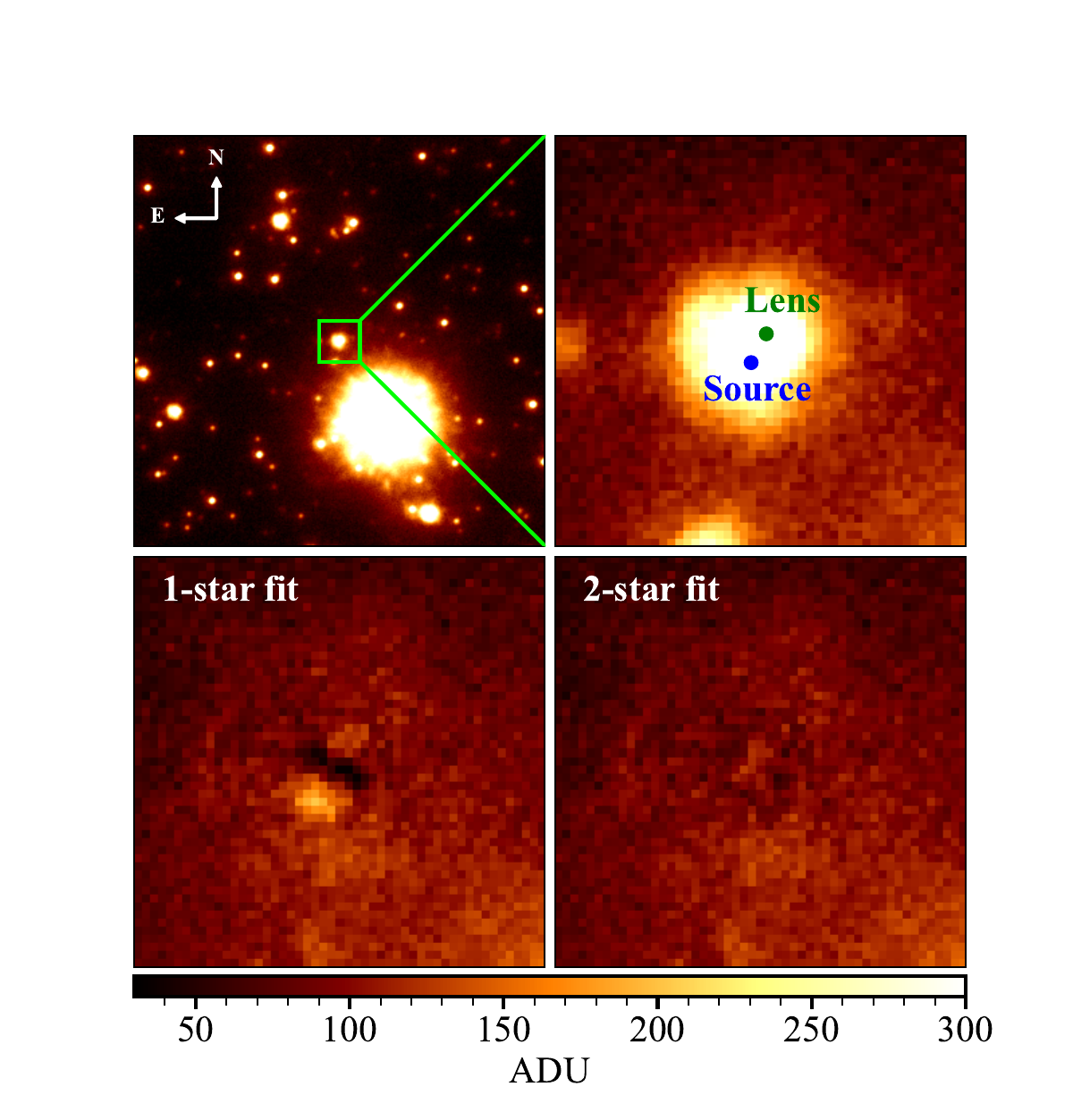}
    \caption{
        \textbf{Top Left:} 
            Co-added image of 6 Keck OSIRIS camera images, with an exposure time of 59.011 seconds each. The target is indicated by the square outline.  
        \textbf{Top Right:} 
            Zoomed image of the OB140676 in which source and lens stars are blended. The source-lens separation in this epoch is $41.22 \pm 1.27$~mas.
        \textbf{Bottom Left:} 
            Residual image of a 1-star PSF fit. a resultant dipole residual indicates that it is not a single star.
        \textbf{Bottom Right:} 
            Residual image for a simultaneous 2-star PSF fit, showing a significantly improved subtraction.
            The color bar represents the pixel intensity in the residual images of the bottom panels.
    }
    \label{fig:quad-panel}
\end{figure*}

We used the \texttt{ALLSTAR} command with the resulting PSF model to perform photometry. 
First, we attempted a single-star fit at the target position and found a dipole feature in the residual image, as shown in the bottom left panel of Figure~\ref{fig:quad-panel}. 
We then performed a two-star fit, which significantly reduced the residuals, as shown in the bottom right panel of Figure~\ref{fig:quad-panel}, yielding an improvement of $\Delta\chi^{2} > 2000$ compared to the single-star fit.
\rev{
We note that allowing a two-star model can partially absorb residual PSF mismatches, leading to modest improvements in $\chi^2$ even for isolated single stars. 
To assess this effect, we performed the same analysis on a nearby isolated star and found an improvement of $\Delta\chi^2 \sim 100$. 
In contrast, the target exhibits a much larger improvement, indicating that the two-star model is physically required rather than an artifact of PSF mismatch.
}
Based on this, it is better to describe this target using a two-star fit rather than one.

To estimate the uncertainties of the measured positions and instrumental magnitudes, we ran an MCMC analysis using the two-star PSF model with the \texttt{emcee} ensemble sampler.
\citet{ter21} first demonstrated fitting in daophot with a \texttt{Fortran} implementation of MCMC.
We find emcee is comparable to the \citet{ter21} performance.
The free parameters in the MCMC analysis are the positions and fluxes of the two stars, $(x_1, y_1, F_1)$ and $(x_2, y_2, F_2)$, a constant background offset within the fitting region, and two additional noise parameters that account for deviations from pure Poisson noise.
The posterior distributions for the relative separations in the east ($\Delta E$) and north ($\Delta N$) directions—defined as the position of Star~1 relative to Star~2— as well as the instrumental magnitudes of Star~1 and Star~2, were used to determine the median values and 68\% credible intervals.
These intermediate results are summarized in Table~\ref{table:Keck_intermediate} under the ``MCMC'' row.
The uncertainties reflect the statistical errors arising from the PSF fitting procedure on the combined image.

To account for additional sources of uncertainty not captured by the MCMC analysis, we also applied the jackknife method \citep{que49, que56, tuk58, tm99}.
While the MCMC method effectively samples the posterior distributions of the model parameters, it does not capture variations in the PSF across individual AO images.
Since the AO correction is imperfect, residual PSF distortions remain and can vary between images, potentially affecting the measuredseparations and instrumental magnitudes.
The jackknife method is a standard approach to estimate the statistical variation arising from such frame-to-frame differences.
For a set of \(N\) dithered images, we create \(N\) stacked images, each constructed from \(N-1\) frames, leaving out a different frame in each stack. 
The parameter \(x\) is measured from each jackknife realization \(x_i\), and the standard error is estimated as  
\begin{equation}
    \mathrm{SE}(x) = \sqrt{\frac{N-1}{N} \sum_{i=1}^{N} \left(x_i - \bar{x}\right)^2},
\end{equation}  
where \(\bar{x}\) is the mean of the \(x_i\).
This jackknife error estimate follows the implementation used in \citet{bha21}.
For our analysis, we generated six jackknife-stacked images from the original six frames.
For each stack, the PSF model was independently constructed using the same selection of PSF stars as for the full-stack image, and the two-star fit was performed on that stack.
The resulting measurements of the relative separations and instrumental magnitudes from each stack were then used to compute the jackknife means and standard errors.
These results are summarized in Table~\ref{table:Keck_intermediate} under the ``Jackknife'' row.

\renewcommand{\arraystretch}{1.3}
\begin{deluxetable*}{lcccc}
    \tablecaption{
        PSF photometry from MCMC and Jackknife Analyses
    }
    \setlength{\tabcolsep}{8pt}
    \tablehead{
        \colhead{} & \colhead{$\Delta E$ (mas)}  & \colhead{$\Delta N$ (mas)} & \colhead{Star~1 (inst.\ mag)} & \colhead{Star~2 (inst.\ mag)}
    }
    \startdata
        MCMC       & $-18.975^{+0.988}_{-1.066}$ & $36.597^{+1.330}_{-1.326}$ & $12.692^{+0.010}_{-0.009}$    & $14.900^{+0.068}_{-0.067}$ \\
        Jackknife  & $-18.685 \pm 0.361$         & $36.029 \pm 0.993$         & $12.714 \pm 0.062$            & $14.925 \pm 0.045$         \\
    \enddata
    \tablecomments{
        The measurements of relative separations and instrumental magnitudes from Keck AO imaging.
        \rev{
        The instrumental magnitudes are defined following the DAOPHOT convention, including an arbitrary zero-point offset.
        }
        The MCMC values are from the full-stack image analysis, while the Jackknife values represent statistical variations from jackknife resampling of the individual frames.
    }
    \label{table:Keck_intermediate}
\end{deluxetable*}

By combining the MCMC and jackknife measurements in quadrature, we derive the final relative proper motion components of the lens-source system 
\rev{in the heliocentric frame} 
as $\mu_{\rm rel,H,E} = -3.02 \pm 0.17$~mas~yr$^{-1}$ and $\mu_{\rm rel,H,N} = 5.83 \pm 0.26$~mas~yr$^{-1}$, as well as the calibrated $K$-band magnitudes of the lens and source, $K_{\rm L} = 16.98 \pm 0.08$ and $K_{\rm S} = 19.19 \pm 0.10$, summarized in Table~\ref{table:Keck_final}.
\rev{
See Section~\ref{subsubsec:relative_proper_motion} for the relation of geocentric and heliocentric proper motions.
}
For the photometric calibration, we used VVV catalog stars whose magnitudes had been calibrated to the 2MASS photometric system.
The heliocentric proper motion components are computed by dividing the measured relative separations by the time baseline of $\Delta t = 6.28$~yr.
Compared to the estimates by \citet{xie21}, who reported the source and lens brightness to be $K_{\rm S} = 19.71 \pm 0.20$, $K_{\rm L} = 16.79 \pm 0.04$, where the source magnitude was adopted from the light-curve analysis of \citet{rat17} and the lens magnitude was inferred as the excess flux in their AO image after subtracting the source contribution, and the geocentric relative proper motion of $\mu_{\rm rel,G} = 3.77 \pm 0.74$~mas~yr$^{-1}$, our heliocentric proper motion measurement is qualitatively larger, although a direct quantitative comparison is not straightforward due to the different reference frames.
Given that the brighter of the two resolved components has a $K$-band magnitude close to the lens brightness reported by \citet{xie21}, and that the fainter component has a magnitude consistent with their reported source brightness (originally derived from the light-curve analysis of \citealt{rat17}), we identify the brighter component as the lens and the fainter component as the source.

\renewcommand{\arraystretch}{1.3}
\begin{deluxetable*}{lcccc}
    \tablecaption{
        Final Results of photometry
    }
    \setlength{\tabcolsep}{8pt}
    \tablehead{
        \colhead{} & \colhead{$\mu_{\rm rel,hel,E}$ (mas\,yr$^{-1}$)} & \colhead{$\mu_{\rm rel,hel,N}$ (mas\,yr$^{-1}$)} & \colhead{Lens (mag)} & \colhead{Source (mag)}
    }
    \startdata
        Combined       & $-3.020 \pm 0.173$                               & $5.826 \pm 0.264$                                & $16.982 \pm 0.080$   & $19.190 \pm 0.095$ \\
    \enddata
    \tablecomments{
        Final results of relative proper motion and calibrated $K_p$-band magnitudes.
        Uncertainties include statistical contributions from the MCMC analysis and systematic contributions from jackknife resampling, added in quadrature.
    }
    \label{table:Keck_final}
\end{deluxetable*}


\section{Inconsistency between Imaging and Light-Curve Constraints}
\label{sec:murel}

\begin{figure*}
    \centering
    \includegraphics[width=0.6\linewidth]{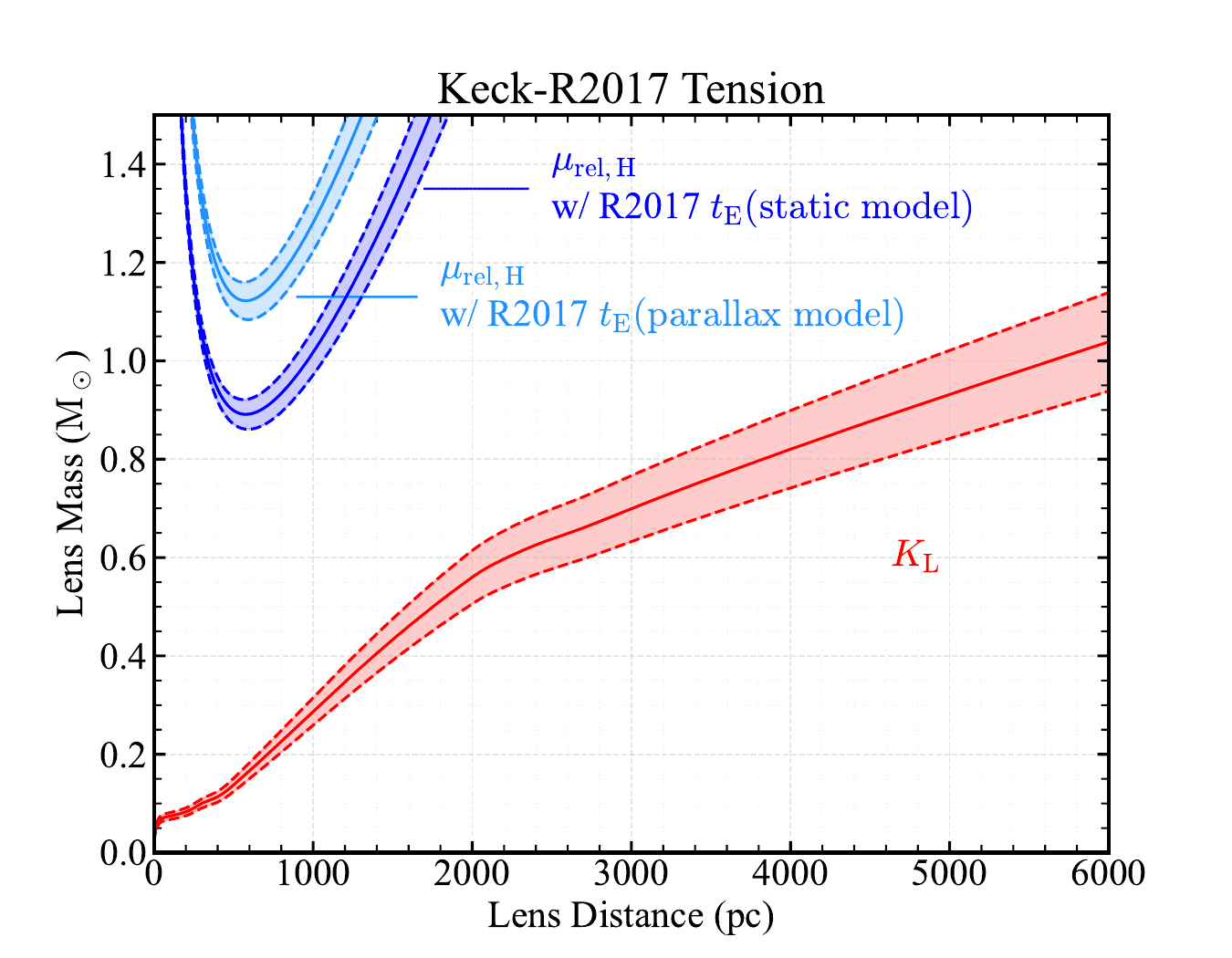}
    \caption{
        The mass-distance relation for OB140676.
        The red curve represents the relation derived from the mass-luminosity relation based on the lens brightness, $K_{\rm L}$ measurement from the Keck image analysis, with a 0.09 mag uncertainty taken into account when converting the brightness to mass.
        The blue curves represent the relation derived from $t_{\rm E}$ from \citet{rat17} and the measured $\mu_{\rm rel}$.
        The deeper blue curve corresponds to the $t_{\rm E}$ from the best-fit model (without parallax), while the lighter blue curve is based on the $t_{\rm E}$ from their \textit{parallax wide} model.
        Note that ``R2017'' on the plot refers to \citet{rat17}.
    }
    \label{fig:mldlplot_first}
\end{figure*}

From the Keck image analysis, we obtained two observables that define the mass-distance relation: the lens brightness in the $K$-band, $K_{\rm L}$, and the lens-source relative proper motion vector measured in the heliocentric frame, $\boldsymbol{\mu}_{\rm rel, H}$.
The apparent lens magnitude $K_{\rm L}$ is given by
\begin{equation}
    K_{\rm L} = {\cal M}_{K, {\rm L}} + 5 \log \left( \frac{D_{\rm L}}{\rm 10~pc} \right) + A_{K, {\rm L}},
\end{equation}
where ${\cal M}_{K, {\rm L}}$ is the absolute magnitude of the lens in $K$-band, and $A_{K, {\rm L}}$ is the extinction at the lens distance $D_{\rm L}$ along the line of sight.
Since the absolute magnitude can be converted to the lens mass $M_{\rm L}$ using a mass-luminosity relation, $K_{\rm L}$ provides a mass-distance relation for the lens.
The red curve in Figure \ref{fig:mldlplot_first} represents this relation, calculated from $K_{\rm L} = 16.98 \pm 0.05$ mag using the same empirical mass-luminosity relation for main-sequence stars that was used in \citet{ben15}, which combines relations from \citet{hen93}, \citet{hen99}, and \citet{del00}.
\rev{
We note that the lens star lies significantly below the red giant branch in the Color Magnitude Diagram (CMD) in Figure 4 of \citet{rat17}, indicating that it is unlikely to be a giant star. 
Therefore, we adopt a mass-luminosity relation appropriate for main-sequence stars.
}
The extinction is estimated following \citet{ben15};
\begin{equation}
    A_{K, {\rm L}} =  \frac{1- e^{-D_{\rm L}/\tau_{\rm dust}}}{1- e^{-D_{\rm rc}/\tau_{\rm dust}}}~A_{K,{\rm rc}},
    \label{eq:extinction}
\end{equation}
where $\tau_{\rm dust} = h_{\rm dust}/\sin{|b|}$ is the dust scale length toward the Galactic bulge at Galactic latitude $b$, and we adopt $h_{\rm dust} = 164~{\rm pc}$ from \citet{nat13}. 
The parameters $D_{\rm rc}$ and $A_{K, {\rm rc}}$ correspond to the mean distance and extinction of the red clump stars in the field, respectively, with values $D_{\rm rc} = 8166 \pm 566$ pc and $A_{K, {\rm rc}} = 0.37\pm0.08$ mag, estimated following \citet{xie21}.
Note that the red-hatched region in Figure~\ref{fig:mldlplot_first} is $95 \%$ uncertainty area that includes all the errors written above.

The lens-source relative proper motion vector, $\boldsymbol{\mu}_{\rm rel, H}$, also constrains a mass-distance relation when combined with the Einstein radius crossing time $t_{\rm E}$;
\begin{equation}
    M_{\rm{L}} = \frac{t_{\rm{E}}^2 \, }{\kappa  \pi_{\rm{rel}}}  \left|\boldsymbol{\mu}_{\rm{rel},\rm{H}} - \frac{\boldsymbol{v}_{\oplus} \pi_{\rm rel}}{\rm AU} \right|^2,
    \label{eq:mu_relation}
\end{equation}
where $\pi_{\rm{rel}} = {\rm AU}\, (D_{\rm L}^{-1} - D_{\rm S}^{-1})$ is the lens-source relative parallax, and $\boldsymbol{\mu}_{\rm{rel},\rm{H}} - \frac{\boldsymbol{v}_{\oplus} \pi_{\rm rel}}{\rm AU}$ represents the lens-source relative proper motion in the geocentric frame. 
The Earth’s transverse velocity at the peak magnification, $\boldsymbol{v}_{\oplus} (N, E) = (0.54, 4.04)$ AU/yr, is used for this conversion.
The blue curves in Figure~\ref{fig:mldlplot_first} show the corresponding mass-distance relation, calculated using the lens-source relative proper motion $(\mu_{\rm rel, HN}, \mu_{\rm rel, HE}) = (5.83, -3.02) \, \pm \, (0.22, 0.17)$ mas/yr measured from the Keck images and the Einstein radius crossing time  $t_{\rm E} = 116.3 \pm 12.2$ days and $t_{\rm E} = 130.5 \pm 12.4$ days for the static and parallax models taken from \citet{rat17}. 
For the plot, we use the source distance $D_{\rm S} = 8.30^{+0.55}_{-0.52}$ kpc, estimated using \texttt{genulens}, a microlensing simulation tool based on a Galactic model \citep{kos21, kosran22}.

As shown in Figure~\ref{fig:mldlplot_first}, the mass--distance relation derived from ${\boldsymbol \mu}_{\rm rel,H}$ and $t_{\rm E}$ exhibits a U-shaped curve.
This feature may be unexpected to some readers, as no such structure appears in the conventional mass--distance relation obtained directly from a measured $\theta_{\rm E}$.  
However, this behavior follows naturally from the form of Equation~(\ref{eq:mu_relation}).
As $D_{\rm L} \rightarrow 0$, the lens-source relative parallax increases without bound ($\pi_{\rm rel} \rightarrow \infty$), and the heliocentric-to-geocentric conversion term becomes dominant. 
Since this term scales linearly with $\pi_{\rm rel}$, the mass estimate in Equation~(\ref{eq:mu_relation}) grows rapidly and eventually diverges. 
Together with the usual divergence as $D_{\rm L} \rightarrow D_{\rm S}$, this produces the characteristic U-shaped profile.

Figure~\ref{fig:mldlplot_first} shows that the two mass–distance relations do not intersect, indicating that no combination of $M_{\rm L}$ and $D_{\rm L}$ can simultaneously satisfy the constraints from the imaging and light-curve analysis.
This inconsistency, therefore, signals a lack of mutual agreement among the key observational inputs used to construct these relations, namely the Einstein timescale $t_{\rm E}$, the heliocentric lens-source relative proper motion $\boldsymbol{\mu}_{\rm rel,H}$, and the lens brightness $K_{\rm L}$. 
Importantly, $\boldsymbol{\mu}_{\rm rel,H}$ and $K_{\rm L}$ were derived jointly from the same Keck imaging analysis and therefore constitute a self-consistent set of constraints, whereas $t_{\rm E}$ was obtained independently from the light-curve modeling.
In light-curve modeling, $t_{\rm E}$ can be susceptible to parameter degeneracies (e.g., with the impact parameter $u_0$) and to biases introduced by systematics in the light curve.
Especially, this event has a relatively faint source star and the light curve wings were poorly covered. In such cases, the $t_{\rm E}$ parameters are sometimes not well determined by ground-based survey data alone \citep{ben24}.

For these reasons, and given the joint consistency of the Keck-derived observables, it is plausible that the previously reported $t_{\rm E}$ contributes substantially to the disagreement shown in Figure~\ref{fig:mldlplot_first}. 
To examine this possibility in more detail, we revisited the light-curve modeling while incorporating the Keck imaging results as additional constraints.
The methodology and outcomes of this reanalysis are described in the following sections.

\section{Light Curve Modeling with Keck Image Constraints}
\label{sec:relight}

In this section, we introduce the constraints derived from our high-angular-resolution image analysis into the light-curve analysis. 
The resulting best-fit light-curve parameters and associated $\chi^2$ values are presented in Table~\ref{tab:lcparams}, while the derived physical parameters, including the relative proper motion, are summarized in Table~\ref{tab:physicalparams}.

\subsection{Light Curve Data}
\label{sec:lightcurve_data}

We use MOA-II, OGLE-IV and Wise data for the image constrained light curve analysis. 
These light curves covered both the anomaly part and baseline of the event.
We use the same data set as \citet{rat17} for OGLE and Wise, but we do not cut the data points, whereas they cut some data points as written in their paper.
Regarding the MOA data set, we applied an improved version of the difference imaging analysis pipeline \citep{bon01} with the detrending method \citep{bon17} to remove systematic errors.
Here the color-dependent effects of differential refraction \citep{ben12} due to the wide MOA-Red filter is a dominant source of the systematics.
Also, the MOA-Red photometry is calibrated to the OGLE-III catalog \citep{szy11} as written below in Eq. \ref{eq:MOA_Red_OGLE_I}.
The number of data points we used for the image- constrained light curve modeling is summarized in Table~\ref{tab:table_lcdata}.

\renewcommand{\arraystretch}{1.3}
\begin{deluxetable*}{lcc}
    \tablecaption{
        The number of data points
    }
    \setlength{\tabcolsep}{12pt}
    \tablehead{
        \colhead{Group} & \colhead{Filter} & \colhead{\# of data points}
    }
    \startdata
        OGLE            & $I$              & 4115  \\
        MOA             & MOA-Red          & 29324 \\
        Wise            & $I$              & 468   \\
    \enddata
    \tablecomments{
        Final data sets used for modeling the planetary microlensing event OB140676.
    }
    \label{tab:table_lcdata}
\end{deluxetable*}

\subsection{Keck Constraints in Light Curve Modeling}
\label{sec:relight_keck}

To incorporate the constraints obtained from the Keck image analysis into our modeling, we introduced an additional chi-square term, $\chi^2_{\rm Keck}$, which penalizes deviations between the model predictions and the Keck-derived values for the source and lens fluxes, as well as the relative proper motion. 
In the following subsections, we describe how the model-predicted values for the source and lens fluxes, as well as the lens–source relative proper motion, are computed from the light curve parameters.

\subsubsection{Source Flux}

The magnitude measured from the Keck image analysis is in the $K$-band, whereas the light curve data provide flux information only in the $I$-band and MOA-Red band.
Thus, the $K$-band source flux cannot be directly determined from the light curve alone.
Therefore, we estimated the $K$-band flux of the source based on the $I_{\rm OGLE_{IV}}$ and $R_{\rm MOA}$ values derived from the light curve modeling.
For this event, the following empirical relations between OGLE-III and MOA magnitudes were adopted:  
\begin{equation}
    \begin{aligned}
        I_{\rm OGLE_{III}} - R_{\rm MOA} &= (27.1479\pm0.0012) + (-0.1980\pm0.0011)(V_{\rm MOA} - R_{\rm MOA}) \\
        V_{\rm OGLE_{III}} - R_{\rm MOA} &= (28.4456\pm0.0012) + (-0.0984\pm0.0011)(V_{\rm MOA} - R_{\rm MOA})
    \label{eq:MOA_Red_OGLE_I}
    \end{aligned}
\end{equation}
We constructed this relation by selecting isolated stars located within $2'$ of the target position, using the OGLE-III photometric maps \citep{szy11}.
By combining these relations with the OGLE-III to OGLE-IV transformation given in \citet{rat17}, we obtained the following conversion formula:  
\begin{equation}
    I_{\rm OGLE_{III}} = (1.038\pm0.010)I_{\rm OGLE_{IV}} + (-0.038\pm0.010)R_{\rm MOA} + (-1.00\pm0.29)
\end{equation}
Using this relation and the source brightness in $I_{\rm OGLE_{IV}}$ and $R_{\rm MOA}$ from the light curve modeling, we first calculated the OGLE-III $I$-band magnitude of the source.
Next, we derived its absolute $I$-band magnitude using the apparent $I$-band magnitude obtained above, the source distance $D_{\rm S}$, and the extinction toward the source $A_{I, \rm S}$, which was derived with Eq. \ref{eq:extinction} but the subscripts $K$ and $\rm L$ were replaced with $I$ and $\rm S$ in the case where $D_{\rm S} < D_{\rm rc}$.
Otherwise, $A_{I, \rm S} = A_{I, \rm rc}$ was applied.
We applied $A_{I,{\rm rc}} = 2.50 \pm 0.05$ from \citet{rat17}. 
Here, $D_{\rm S}$ was included as an additional fitting parameter in the MCMC analysis, appended to the standard light curve parameter set $\theta_{\rm lc} = (t_{0}, t_{E}, u_{0}, q, s, \alpha, \rho, \pi_{\rm EE}, \pi_{\rm EN})$, while $D_{\rm rc}$ and $A_{I, \rm rc}$ were randomly sampled from Gaussian distributions.
The prior distribution for $D_{\rm S}$ was constructed using a Galactic model implemented via \texttt{genulens} \citep{kosran22}, and calibrated to be consistent with the observational constraints from the Keck image analysis, as described in Section~\ref{sec:murel}.
\rev{The $D_{\rm S}$ prior is $8.2\pm0.6\, \rm kpc$, whose uncertainty is the dispersion of simulated source distances for detectable microlensing events under the adopted Galactic model.}
Finally, the apparent $K$-band magnitude of the source was computed from its absolute $K$-band magnitude, derived from the $I$-band magnitude via the empirical main-sequence mass-luminosity relation \citep{ben15}, the source distance $D_{\rm S}$, and the $K$-band extinction, calculated from $A_{I, \rm S}$ using the extinction law of \citet{WC19}.

\subsubsection{Lens Flux}

To calculate the $K$-band magnitude of the lens, we proceeded as follows. 
First, we computed the angular radius of the source star, $\theta_*$, using the $I$- and $K$-band magnitudes of the source and the empirical relation given by \citet{boy14}. 
Then, we calculated the angular Einstein radius $\theta_{\rm E}$ using $\theta_{\rm E} = \theta_*/\rho$, where $\rho$ is the normalized source radius parameter from the light curve fit.
Next, we derived the microlens parallax $\pi_{\rm E}$ from the light curve parameters as $\pi_{\rm E} = \sqrt{\pi_{\rm EE}^2 + \pi_{\rm EN}^2}$, and then calculated the lens mass using the standard relation:  
\begin{equation}
    M_{\rm L} = \frac{\theta_{\rm E}}{\kappa \pi_{\rm E}}.
\end{equation}
With the estimated lens mass, we used the empirical mass-luminosity relation for main-sequence stars from \citet{ben15}, which combines relations from \citet{hen93, del00, hen99} for different stellar mass ranges with smooth interpolation between them, to derive the absolute $K$-band magnitude of the lens under the assumption of a main-sequence lens star.
To convert this into an apparent magnitude, we calculated the lens distance $D_{\rm L}$ using
\begin{equation}
    D_{\rm L} = \left( \frac{\theta_{\rm E} \pi_{\rm E}}{\rm AU }+ \frac{1}{D_{\rm S}} \right)^{-1}.
\end{equation}
Finally, we applied the $K$-band extinction correction estimated using the method described in Section \ref{sec:murel}, and obtained the apparent $K$-band magnitude of the lens.

\subsubsection{Relative Proper Motion}
\label{subsubsec:relative_proper_motion}

Since the light-curve analysis is performed in the geocentric reference frame, while our Keck AO measurements provide the relative proper motion in the heliocentric frame, we convert the heliocentric proper motion components to the geocentric frame in order to enable a consistent comparison.
The heliocentric relative proper motion between the lens and source, ${\boldsymbol\mu}_{\rm rel, H}$, was calculated in both the North and East directions via the following equation \citep{don09}:
\begin{equation}
    {\boldsymbol\mu}_{\rm rel,G} = {\boldsymbol\mu}_{\rm rel,H} - \frac{{\boldsymbol v}_{\oplus}\pi_{\rm rel}}{\rm AU}
    \label{eq:murelG}
\end{equation}
This calculation requires the geocentric proper motion ${\boldsymbol\mu}_{\rm rel, G}$, the Earth’s projected velocity ${\boldsymbol v}_{\oplus}$ at the time of the event, and the relative parallax $\pi_{\rm rel}$. 
The Earth’s projected velocity is fixed to the value at the time of peak magnification, while ${\boldsymbol\mu}_{\rm rel, G}$ and $\pi_{\rm rel}$ are derived from the light curve parameters.
The posterior distributions of the geocentric relative proper motion obtained from this conversion are summarized in Table~\ref{tab:physicalparams}.

\subsubsection[Comparison of chi2 values]{Comparison of $\chi^2$ values}
To evaluate the consistency between the light curve model and the constraints derived from the Keck image analysis, we computed a penalty chi-square term, $\chi^2_{\rm Keck}$, at each MCMC step. 
This term quantifies the deviation between the observational values $\boldsymbol{p}_{\rm Keck} = (K_{\rm S}, K_{\rm L}, \mu_{\rm rel,HN}, \mu_{\rm rel,HE})$ and the model-predicted values $\boldsymbol{p}_{\rm lc}$ computed from the light curve parameters, according to the following expression: 
\begin{equation}
    \chi^{2}_{\rm Keck} = \sum_{i} \frac{(p_{i, \rm lc} - p_{i, \rm Keck})^2}{\sigma_{i, \rm Keck}^2}
    \label{eq:chi2_keck}
\end{equation}
The total chi-square used in the MCMC likelihood is given by the sum of the light-curve and Keck AO constraints:
\begin{equation}
    \chi^2_{\rm total} = \chi^2_{\rm lc} + \chi^2_{\rm Keck}.
\end{equation}
The individual contributions from $\chi^2_{\rm lc}$ and $\chi^2_{\rm Keck}$, as well as the total $\chi^2$, are listed in Table~\ref{tab:lcparams}.


\section{Final Result of light curve analysis with Keck Constraints}
\label{sec:result}

\begin{figure}
    \centering
    \includegraphics[width=1.0\linewidth]{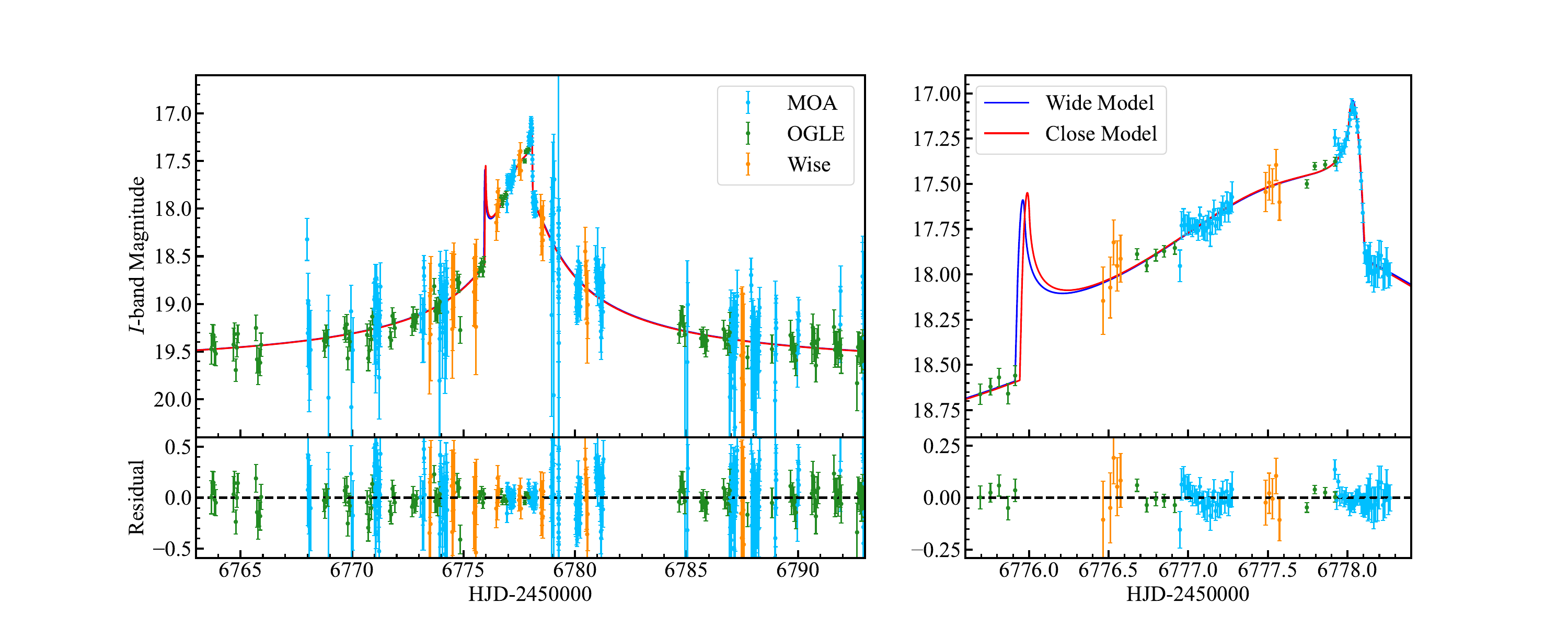}
    \caption{
        The light curve data for OB140676 from MOA, OGLE and Wise.
        The best fit close and wide models are plotted in the red and blue lines, respectively. The bottom panels show the residual from the close model. The right panel is the zoom-in around the planetary anomaly.
    }
    \label{fig:lc_result}
\end{figure}

\begin{figure}
    \centering
    \includegraphics[width=0.9\linewidth]{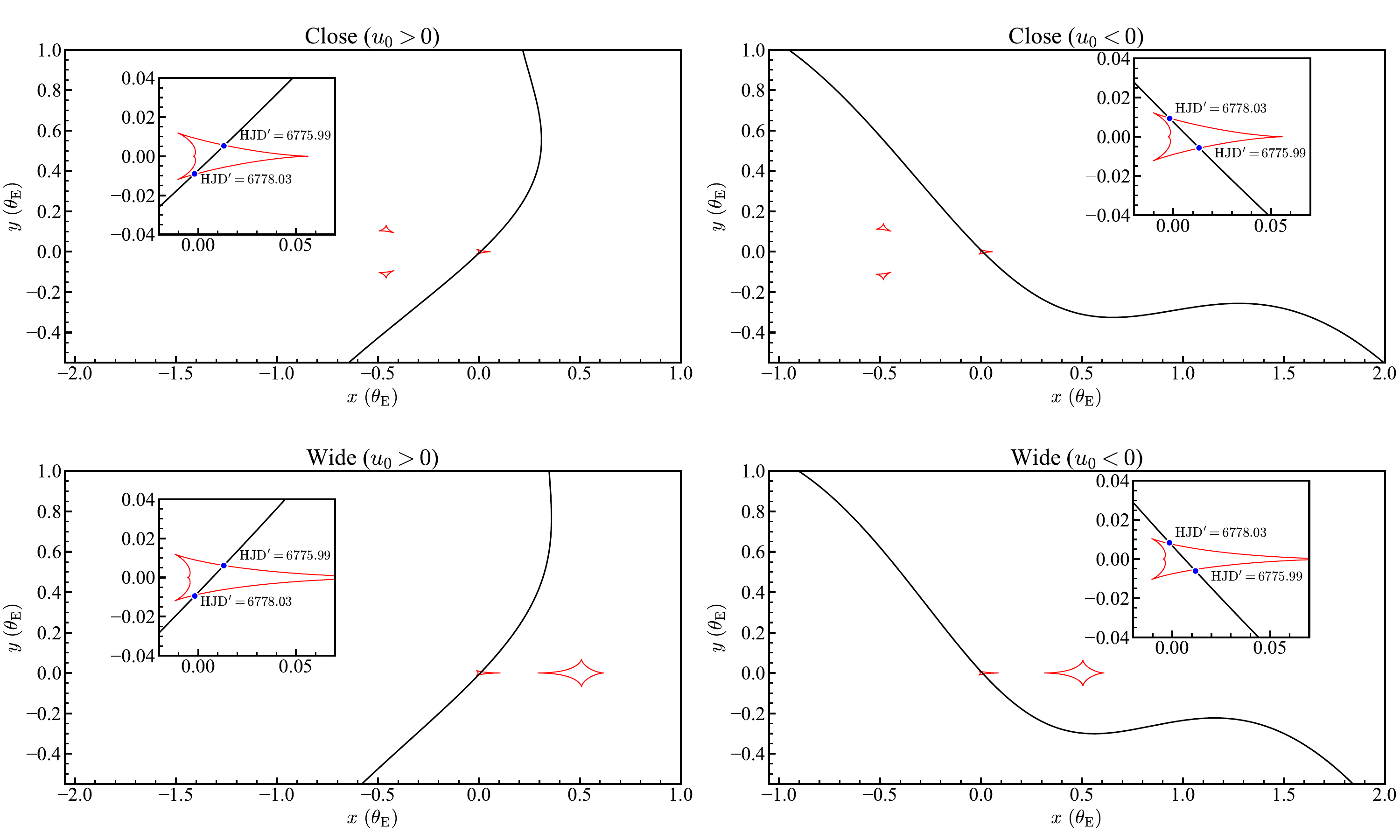}
    \caption{
        The caustics and source trajectories \rev{for the close (top) and wide (bottom) models} with $\pm u_{0}$ solutions. The inset shows a magnified view around the central caustic, highlighting the two caustic crossings.
    }
    \label{fig:caustic}
\end{figure}

\begin{figure*}
    \centering
    \includegraphics[width=0.6\linewidth]{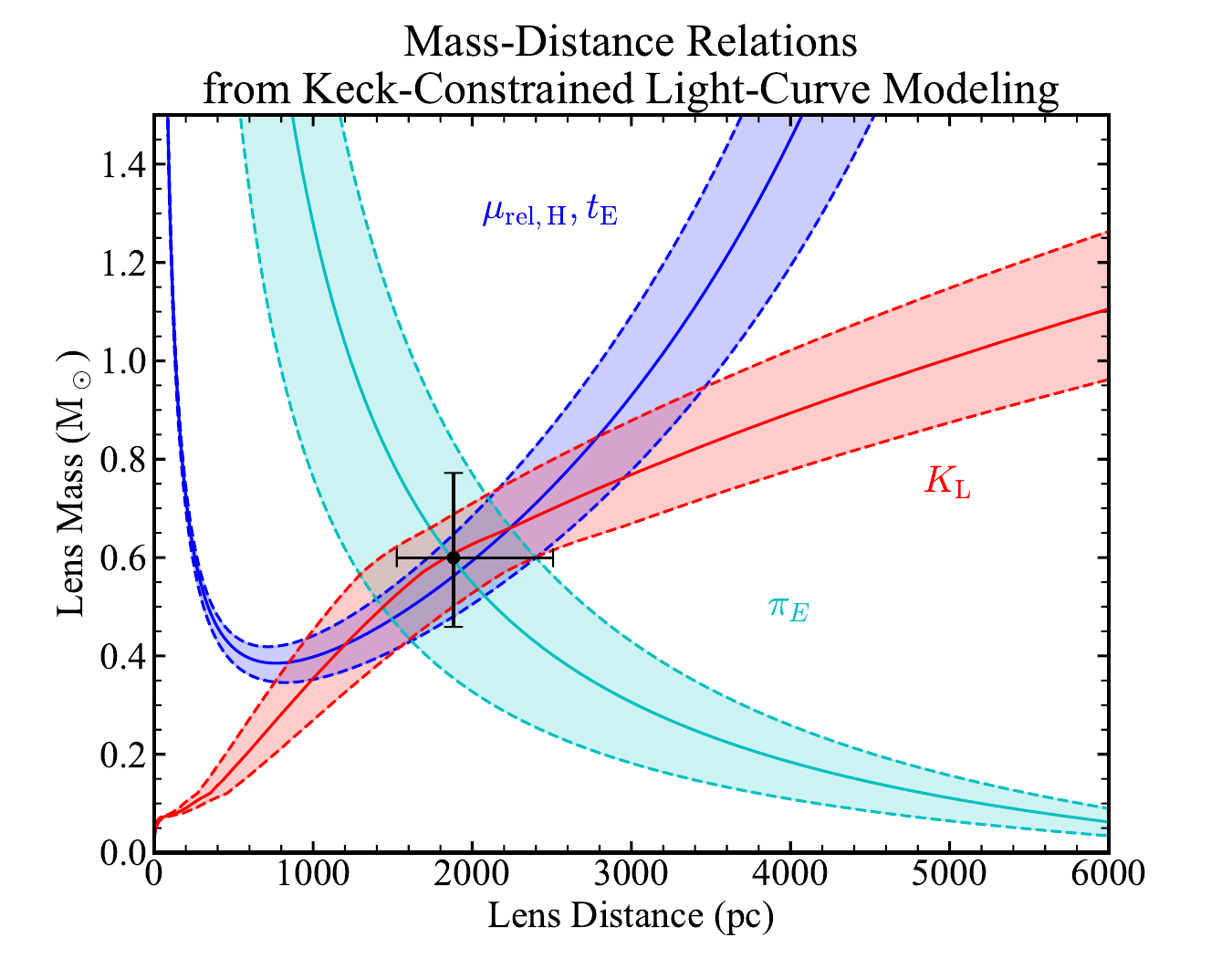}
    \caption{
        Lens mass–distance relation computed from the \rev{combined weighted} posterior distributions of our Keck-constrained MCMC light-curve analysis.  
        The red curve shows the constraint from the lens brightness, the blue curve represents the constraint based on the Einstein timescale $t_{\rm E}$ combined with the heliocentric relative proper motion ${\boldsymbol \mu}_{\rm rel,H}$, and the cyan curve corresponds to the constraint from the microlensing parallax vector ${\boldsymbol \pi}_{\rm E}$.  
        The shaded regions denote the corresponding 68\% intervals.
    }
    \label{fig:mldlplot_final}
\end{figure*}

Using the Keck AO constraints described in Section~\ref{sec:relight_keck}, we performed a grid search across the binary-lens parameter space, incorporating the Keck-derived priors. 
The grid search was performed over $(q, s, \alpha)$.
From the resulting models, we found the best-fitting solutions for both the close ($s < 1$) and wide ($s > 1$) models. 
\rev{We also found the degeneracy between the models with $u_{0} < 0$ and $u_{0} > 0$ in each close and wide solution. 
The $\pm u_{0}$ models have almost same parameter values except for $\alpha$ that shows opposite sign. 
The chi-square difference between the $\pm u_{0}$ solutions is less than 1. 
As for the comparison of the close and wide models, the chi-square difference is $\Delta\chi^2 = 2.499$, 
}
indicating that both models provide comparably good fits to the data. 
We then carried out full MCMC analyses for these models under the same Keck constraints.
The best fit microlensing parameters for the close and wide models \rev{with $\pm u_{0}$} are summarized in Table~\ref{tab:lcparams}.
\rev{We also show the MCMC averages for each parameter in the table.}
The light curve and caustic geometry for the best-fit close \rev{and wide} models are shown in Figures~\ref{fig:lc_result} and~\ref{fig:caustic}, respectively.
The resultant median values and $1\sigma$ uncertainties of the physical parameters are summarized in Table~\ref{tab:physicalparams}\rev{, where we combined the degenerated four models by applying the $e^{-\Delta \chi^2/2}$ weight }. 
We show the lens mass-distance relations for the combined weighted best fit model in Figure \ref{fig:mldlplot_final}.

\medskip
\noindent

\rev{
The lens host star has a mass of $M_{\rm host} = 0.60^{+0.17}_{-0.14}\,M_\odot$, and the planet has a mass of $m_{\rm p} = 3.11^{+1.11}_{-0.63}\,M_{\rm J}$.
The lens distance is $D_{\rm L} = 1.88^{+0.63}_{-0.35}$~kpc, and the projected planet--host separation is $a_\perp = 2.04^{+0.44}_{-0.35}$~au and $a_\perp = 3.72^{+0.92}_{-0.72}$~au for the close and wide model, respectively.
}

\renewcommand{\arraystretch}{1.3}
\begin{deluxetable*}{lcccccccc}
    \tabletypesize{\scriptsize}
    \tablecaption{
        Best-Fit Microlensing Parameters
    }
    \setlength{\tabcolsep}{6pt}
    \tablehead{
        \colhead{}            & \multicolumn{2}{c}{R2017}                                & \colhead{} & \multicolumn{5}{c}{This Work}                               \\
        \colhead{Parameter}   & \colhead{Close w/ parallax} & \colhead{Wide w/ parallax} & \colhead{} & \colhead{Close,$u_0<0$} & \colhead{Close,$u_0>0$} & \colhead{Wide,$u_0<0$} & \colhead{Wide,$u_0>0$} & \colhead{MCMC Averages} \\
    }
    \startdata
        $t_0$ (HJD')          & $6777.309 \pm 0.052$ & $6777.319 \pm 0.045$ && $6777.269$  & $6777.282$  & $6777.299$  &$6777.270$   & $6777.3217^{+0.0403}_{-0.0358}$\\
        $t_{\rm E}$ (days)    & $126.4 \pm 10.5$     & $130.5 \pm 12.4$     && $95.8$      & $97.1$      & $105.8$     &$94.7$       & $88.9^{+10.7}_{-10.4}$     \\
        $u_0$ ($10^{-3}$)     & $3.9 \pm 0.4$        & $3.6 \pm 0.4$        && $-5.3$      & $5.3$       & $-4.6$      &$5.3$        & $5.48^{+0.79}_{-0.67}$     \\
                              &                      &                      &&             &             &             &             & $-5.36^{+0.64}_{-0.81}$    \\
        $q$ ($10^{-3}$)       & $4.31 \pm 0.07$      & $4.60 \pm 0.37$      && $4.87$      & $4.44$      & $4.45$      &$5.15$       & $5.17^{+0.78}_{-0.65}$     \\
        $s$                   & $0.757 \pm 0.013$    & $1.358 \pm 0.021$    && $0.785$     & $0.794$     & $1.284$     &$1.287$      & $0.786^{+0.011}_{-0.011}$  \\
                              &                      &                      &&             &             &             &             & $1.299^{+0.019}_{-0.019}$  \\
        $\alpha$ (rad)        & $2.284 \pm 0.046$    & $2.229 \pm 0.030$    && $3.927$     & $-3.902$    & $-2.314$    &$2.333$      & $3.969^{+0.057}_{-0.047}$  \\
                              &                     &                       &&             &             &             &             & $-3.972^{+0.051}_{-0.055}$ \\
                              &                     &                       &&             &             &             &             & $-2.274^{+0.052}_{-0.042}$ \\
                              &                     &                       &&             &             &             &             & $2.265^{+0.048}_{-0.053}$  \\
        $\rho$ ($10^{-4}$)    & $2.43 \pm 0.27$      & $2.61 \pm 0.27$      && $2.62$      & $2.446$     & $2.55$      &$2.73$       & $2.96^{+0.46}_{-0.36}$     \\
        $\pi_{\rm EN}$        & $-2.24 \pm 0.51$     & $-1.81 \pm 0.22$     && $0.213$     & $0.227$     & $0.204$     &$0.201$      & $0.206^{+0.049}_{-0.047}$  \\
        $\pi_{\rm EE}$        & $0.57 \pm 0.48$      & $0.06 \pm 0.18$      && $-0.218$    & $-0.246$    & $-0.211$    &$-0.190$     & $-0.205^{+0.066}_{-0.068}$ \\
        $D_{\rm S}$ (kpc)     & --                   & --                   && $8.15$      & $8.32$      & $8.02$      &$8.01$       & $8.07^{+0.48}_{-0.58}$     \\
        \hline
        source mag.           & --                   & --                   && $23.297$    & $23.319$    & $23.400$    & $23.274$    & $23.215^{+0.129}_{-0.143}$ \\
        blend mag.            & --                   & --                   && $19.822$    & $19.822$    & $19.819$    & $19.822$    & $19.824^{+0.006}_{-0.004}$ \\
        baseline mag.         & --                   & --                   && $19.780$    & $19.781$    & $19.781$    & $19.780$    & $19.779^{+0.001}_{-0.001}$ \\
        \hline
        $\chi^2_{\rm lc}$     & --                   & --                   && $33903.448$ & $33902.639$ & $33899.338$ & $33905.371$ & -- \\
        $\chi^2_{\rm Keck}$   & --                   & --                   && $27.417$    & $28.254$    & $34.026$    & $28.460$    & -- \\
        d.o.f.                & --                   & --                   && $33891$     & $33891$     & $33891$     & $33891$     & -- \\
        $\chi^2$/d.o.f.       & --                   & --                   && $1.00118$   & $1.00118$   & $1.00125$   & $1.00126$   & -- \\
    \enddata
    \tablecomments{
        Best-fit microlensing parameters for the close and wide binary-lens models including parallax effects.
        \citet{rat17} values (first two columns,  noted as "R2017”) are listed for comparison.
        The $\chi^2$ values of R2017 are omitted because they are not directly comparable due to different datasets and modeling procedures.
        The values labeled ``This Work'' are quoted as the median with uncertainties corresponding to $^{84{\rm th}}_{16{\rm th}}$ percentiles, i.e., $50{\rm th}^{84{\rm th}-50{\rm th}}_{16{\rm th}-50{\rm th}}$.
        \rev{The source, blend and baseline magnitudes are listed in $I$-band.}
    }
    \label{tab:lcparams}
\end{deluxetable*}

\renewcommand{\arraystretch}{1.3}
\begin{deluxetable*}{lcccc}
    \tablecaption{
        Physical Parameters for OGLE-2014-BLG-0676
    }
    \setlength{\tabcolsep}{8pt}
    \tablehead{
        \colhead{Prameter} & \colhead{Units}  & \colhead{16th} & \colhead{50th} & \colhead{84th}
    }
    \startdata
        Host star mass, $M_{\rm host}$                          & $M_{\odot}$     & 0.46 & 0.60 & 0.77 \\
        Planet mass, $m_{\rm p}$                                & $M_{\rm J}$     & 2.48 & 3.11 & 4.22 \\
        Lens distance, $D_{\rm L}$                              & kpc             & 1.53 & 1.88 & 2.51 \\
        Projected separation (close), $a_{\perp}$               & au              & 1.69 & 2.04 & 2.48 \\
        Projected separation (wide), $a_{\perp}$                & au              & 3.00 & 3.72 & 4.64 \\
        Angular Einstein radius, $\theta_{\rm E}$               & mas             & 1.24 & 1.38 & 1.53 \\
        Relative proper motion (geocentric), $\mu_{\rm rel,G}$  & mas\,yr$^{-1}$  & 5.16 & 5.65 & 6.24 \\
    \enddata
    \tablecomments{
        16th, 50th (median), and 84th percentiles of the combined posterior distributions of the physical parameters \rev{by applying the $e^{-\Delta \chi^2/2}$ weight.}
    }
    \label{tab:physicalparams}
\end{deluxetable*}

\section{Discussion}
\label{sec:discus}

\subsection{\rev{Context of the Planet in the Microlensing Population}}
\rev{Our analysis shows that the planet, OGLE-2014-BLG-0676b, has a mass of $3 M_{\rm Jup}$ orbiting an early M or K-type host star with a mass of $0.6M_\odot$.
Although the host star in this event is confirmed to be a main sequence star from the detection of lens flux, a $\sim 0.6M_\odot$ lens identified only from microlensing light curve information can still have a substantial probability of being a white dwarf. 
It is still unclear how frequently white dwarfs host planets compared to main-sequence stars. 
So far, two planetary systems around white dwarfs have been discovered through microlensing observations \citep{bla21, zhang24WD}. 
High-spatial-resolution imaging is therefore required to distinguish whether the lens is a main-sequence star or a white dwarf, even when the lens mass is constrained from light-curve modeling. 
Hence, the joint analysis of microlensing light curves and high-resolution imaging data, as performed for this event, will be important for future statistical studies of the frequency of planets around white dwarfs. 
In this context, this event also provides an important sample of a planetary system with a $\sim 0.6 M_{\odot}$ main sequence star.}

\subsection{Comparison with Previous Studies}

\begin{figure*}
    \centering
    \includegraphics[width=1.0\linewidth]{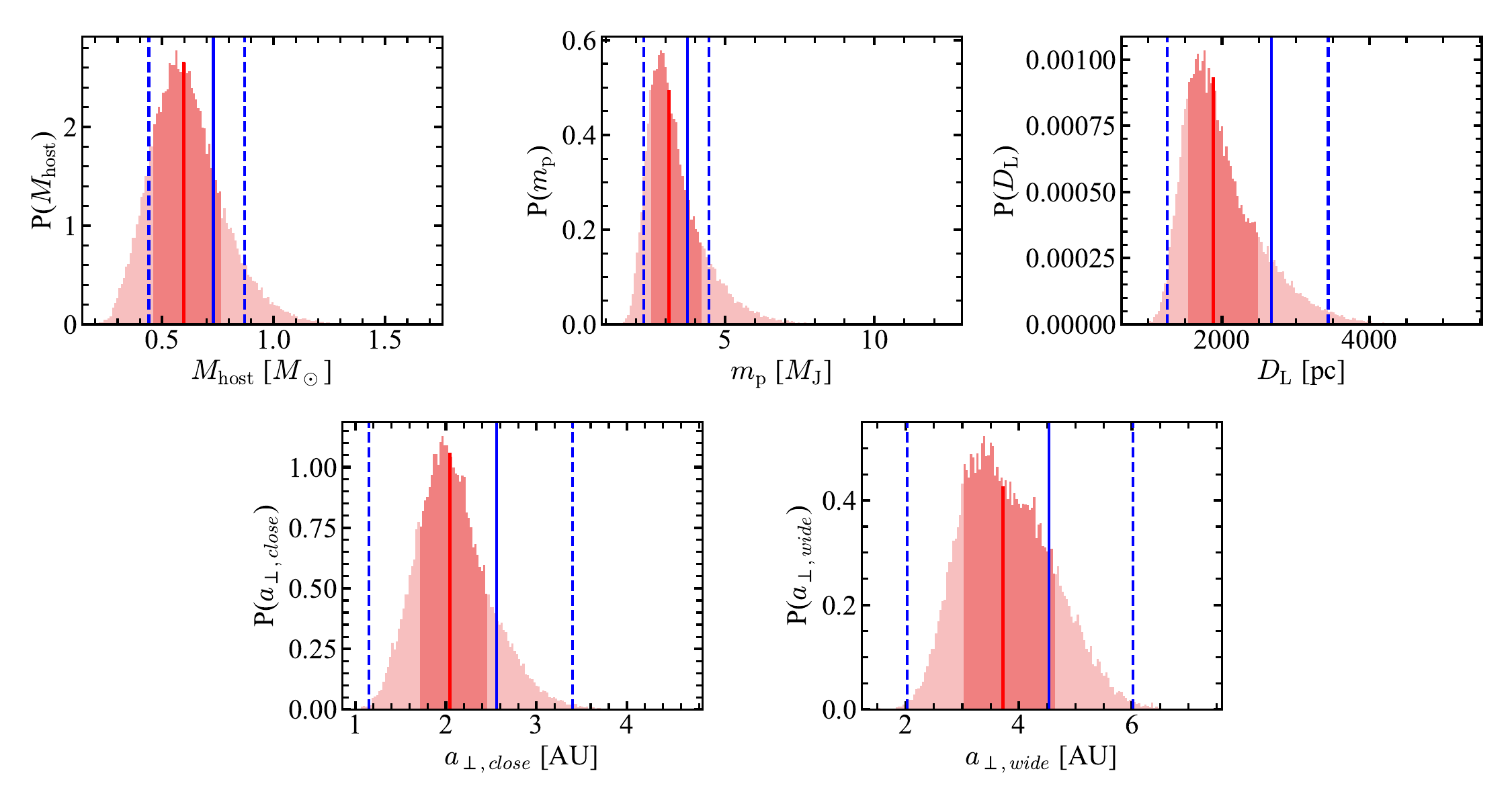}
    \caption{
        The posterior probability distributions for the host mass ($M_{\rm host}$), the planetary companion mass ($m_{\rm p}$), their separation ($a_{\perp}$), and the distance to the lens system ($D_{\rm L}$) are shown. 
        The red histograms represent the \rev{combined weighted} posterior distributions obtained in this work using light-curve modeling with Keck imaging constraints. 
        The deep red shaded region corresponds to the $\pm 1\sigma$ credible interval of our posterior estimate, while the solid red vertical line marks the median value. 
        The blue solid and dashed lines indicate the median (solid) and $\pm 1\sigma$ (dashed) ranges, respectively, obtained by \citet{xie21}.
    }
    \label{fig:mldl_distribution}
\end{figure*}

Figure~\ref{fig:mldl_distribution} shows the posterior distributions obtained from our Keck AO--constrained light-curve modeling.
The blue solid and dashed vertical lines indicate the median and $1\sigma$ ranges from \citet{xie21}, respectively, as a comparison.
The figure highlights that our analysis yields substantially tighter constraints on the key parameters of $M_{\rm L}$, $M_{\rm p}$, $D_{\rm L}$, and $a_{\perp}$.
These improvements arise from incorporating the directly measured lens and source flux, as well as lens--source separation from Keck AO image into the light-curve modeling, enabling us to obtain physically consistent solutions.

The best-fit Keck AO--constrained light-curve models obtained in this work differ from those reported by \citet{rat17}, particularly in the parallax parameters and the Einstein timescale.
Their light-curve–only solution yielded unusually large parallax components and a long Einstein timescale of $100 - 130$ days.
In contrast, our Keck AO--constrained modeling prefers physically plausible parallax values and a significantly shorter timescale of $t_{\rm E} = 90.2 \pm 10.1$ days.
Such differences are not surprising: for long-timescale events, $t_{\rm E}$ can remain weakly constrained when the coverage of the rising and falling wings of the light-curve is limited, because the degeneracy between $t_{\rm E}$ and the source flux becomes stronger in this regime \citep{ben24}.
By incorporating both the lens--source separation and the source brightness measured from the Keck images, our modeling breaks this degeneracy and converges on a more robust solution.

The key methodological difference between this work and \citet{xie21} lies in whether the high-resolution imaging information is incorporated directly into the light-curve modeling.
Because their AO data did not resolve the lens and source, \citet{xie21} adopted the source flux and $\theta_{\rm E}$ values from \citet{rat17} and attributed the excess flux in their AO images to the lens.
However, the underlying light-curve solution from \citet{rat17} is the one that we now find to be disfavored once the Keck AO--constraints are applied.
In other words, the physical parameters inferred by \citet{xie21} rely on a light-curve model that is not fully consistent with the high-resolution imaging.
By contrast, our analysis integrates the Keck measurements of the lens--source separation and source flux and lens flux directly into the MCMC light-curve modeling, ensuring that the derived lens and planet parameters simultaneously satisfy both the imaging data and the photometric data.
This Keck AO--constrained light-curve modeling approach is therefore essential for obtaining a self-consistent and physically reliable solution.

\subsection{Reliability of Our Measurements}

The reliability of our measurements is supported by the fact that the lens and source are directly resolved in the Keck AO image.
The measured separation of $41.2 \pm 1.7$ mas immediately yields a heliocentric proper motion of $\mu_{\mathrm{rel,H}} \sim 6~\mathrm{mas~yr^{-1}}$, a value that is larger than those inferred from light-curve--only analyses ($\mu_{\mathrm{rel,G}} \sim 4~\mathrm{mas~yr^{-1}}$; \citealt{rat17}).
Similarly, the source brightness measured from the Keck images
($K_s = 19.19 \pm 0.10$ mag) is about 0.5 mag brighter than the estimate from previous studies.
These direct observational measurements play a central role in guiding the light-curve modeling toward a consistent physical solution.

Our analysis adopts a main-sequence lens and applies the corresponding mass–luminosity relation.
The observed lens flux ($K_L = 16.98 \pm 0.08$ mag) agrees well with this assumption, while compact remnants would be substantially fainter.
While a white-dwarf lens cannot be entirely excluded, the observed brightness makes this scenario highly unlikely.
We also adopt the extinction values from \citet{xie21}, and the propagated uncertainty ($A_I = 2.50 \pm 0.05$ mag) has only a minor influence on the lens mass determination.

Another potential caveat is that the two objects detected in the Keck AO image may not correspond to the true source and lens, but could instead be chance-aligned field stars or bound companions.
In particular, if the true source were located within $15$--$20$~mas of the lens, corresponding to $\mu_{\rm rel} \lesssim 3~\mathrm{mas~yr^{-1}}$, the solutions of previous studies in terms of $t_{\rm E}$ and source brightness could remain viable.
While we cannot fully rule out this possibility, a simple probability estimate based on the field star density $\rho \approx 0.18~\mathrm{arcsec^{-2}}$ derived from high-resolution Magellan imaging \citep{xie21} indicates that the chance of a random background star appearing within 20~mas of the lens is extremely small ($\sim 2 \times 10^{-4}$).
Therefore, a more plausible concern is the presence of a bound companion to the source or lens. 
The existence of a bound companion cannot be fully ruled out; however, our three-star fitting tests reveal no signs of an extra object in the Keck images.

Overall, the Keck AO measurements provide a strong empirical foundation for the derived physical parameters.
The remaining uncertainties—such as the exact nature of the lens or the possibility of a bound companion—represent natural limitations of the current data rather than inconsistencies in the modeling.
These considerations also point to promising opportunities for future work: additional high-resolution imaging could help address the remaining uncertainties and potentially strengthen the current scenario.

\subsection{Confirmation by expected Roman data}
\label{sec:roman}

OB140676 is located outside of the planned \textit{Roman}'s Galactic Bulge Time Domain Survey fields \citep{spe15, ROTAC, ter25}. However, the Wide Field Imaging Survey in Roman's Galactic Plane Survey \citep{GPS} will cover the location of this event with F129(``$J$"), F158(``$H$") and F213 (``$K$") filters.
The recommended survey program will image the field with ``$J$" and ``$K$" filters around March 2027 and with ``$H$" filter around March 2028.
The expected magnitudes of the lens and source and separation \rev{derived from the combined weighted posterior distribution} are summarized in Table \ref{tab:roman}. 
The magnitudes of the lens and source in each band are within the measurable range. Considering the expected PFS sizes with these filters, the lens and source will be well resolved with the planned 60 second integration.
By adding another one or two epochs of the high-spatial imaging, the {\it Roman} data will confirm our Keck image constrained light curve modeling.
Moreover, using several filters allows us to use several different dust extinction values and mass-luminosity relations, leading to the mass-distance relations depending on the used filters.
Hence, more robust measurement of the physical parameters on the lens and source will be expected.


\renewcommand{\arraystretch}{1.4}
\begin{deluxetable*}{lccc|lccc}
    \tablecaption{
        Expected source and lens magnitudes and predicted separations for \textit{Roman} observations
    }
    \tablehead{
        \colhead{\shortstack{\textbf{Photometry}\\\textbf{(Vega mag)}}} &
        \colhead{16th} & \colhead{50th} & \colhead{84th} &
        \colhead{\shortstack{\textbf{Astrometry}\\\textbf{(Source $\rightarrow$ Lens; mas)}}} &
        \colhead{16th} & \colhead{50th} & \colhead{84th}
    }
    \startdata
        \textit{Source} &  &  &  & \textit{2027 Mar} &  &  \\
        $J$ & 20.58 & 21.00 & 21.34 & East  & -34.17 & -29.20 & -25.36 \\
        $H$ & 19.56 & 19.93 & 20.24 & North &  49.09 &  54.60 &  61.18 \\
        $K$ & 19.38 & 19.49 & 19.59 &  &  &  \\
        \hline
        \textit{Lens} &  &  &  & \textit{2028 Mar} &  &  \\
        $J$ & 16.49 & 17.82 & 19.47 & East  & -36.83 & -31.47 & -27.34 \\
        $H$ & 15.77 & 16.98 & 18.65 & North &  52.91 &  58.85 &  65.95 \\
        $K$ & 16.11 & 16.66 & 17.24 &  &  &  \\
    \enddata
    \tablecomments{
        Photometric magnitudes are given in the Vega system.
        Astrometric separations are expressed in milliarcseconds (mas) and are defined as the vector from the source to the lens in East and North coordinates.
        The listed values represent the 16th, 50th (median), and 84th percentiles of the predicted distributions.
    }
    \label{tab:roman}
\end{deluxetable*}


\section{Conclusion}
\label{sec:conclusion}

In this work, we combined Keck AO imaging obtained 6.3 years after the microlensing event OGLE-2014-BLG-0676 with light curve modeling to derive accurate and consistent physical parameters of the lens system. 
The high-resolution imaging allowed us to directly resolve the lens and source, thereby eliminating the blended-light assumption adopted in the previous study. 
Incorporating these observational constraints into the modeling revealed that the longer-timescale solution favored in earlier analyses was biased by possible systematics in the light curve data, and instead supported a shorter-timescale, physically consistent solution.

We determined the lens mass, distance, and planet mass to be 
$M_{\rm host} = 0.60^{+0.17}_{-0.14}\,M_{\odot}$, 
$D_{\rm L} = 1.88^{+0.63}_{-0.35}$ kpc, and 
$m_{\rm p} = 3.11^{+1.11}_{-0.63}\,M_{\rm J}$, 
respectively, with a projected separation of 
$a_{\perp} = 2.04^{+0.44}_{-0.35}$ au and $a_{\perp} = 3.72^{+0.92}_{-0.72}$ au for the close and wide model. 
These results show that image constrain light curve modeling can significantly reduce the uncertainties in the physical parameters, highlighting the importance of combining high-resolution imaging with microlensing light curve modeling.

Our findings demonstrate that high spatial resolution imaging can provide powerful constraints to break degeneracies and mitigate systematic biases in photometric data, thereby enabling robust determinations of lens and planet properties. 
As written in Sec \ref{sec:roman}, the \textit{Roman} Galactic Plane Survey will cover the location of this event. 
The Roman data will confirm our result and make better constraints on the physical parameters of the lens and source stars by the multi epoch imaging with
several filters. 
This method can be applied to the other previously detected events.


\section*{Acknowledgments}
\begin{acknowledgments}
The MOA project is supported by JSPS KAKENHI Grant Number 
JP16H06287, JP22H00153, JP23KK0060 and JP25H00668.
D.S. is supported by JSPS KAKENHI Grant Number JP19KK0082, JP20H04754, JP24H01811 and JPJSCCA20210003.
N.K. is supported by JSPS KAKENHI grant No. 24K17089.
D.P.B., A.B., S.K.T., and A.V. acknowledge support from NASA through grant NASA-80NSSC18K0274.
J.P.B. was supported by the University of Tasmania through the endowed Warren Chair in Astronomy, the Australian Government through the Australian Research Council
Discovery Project Grants 240101842, and the ANR-24-CE31-3263 SPACE-MLENS.
\end{acknowledgments}


\appendix

\section{
\rev{Visualization of the Target Compared to a Single Star}
}
\label{app:single_vs_traget}

\renewcommand{\thefigure}{A\arabic{figure}}
\setcounter{figure}{0}

\begin{figure*}
    \centering
    \includegraphics[width=0.7\linewidth]{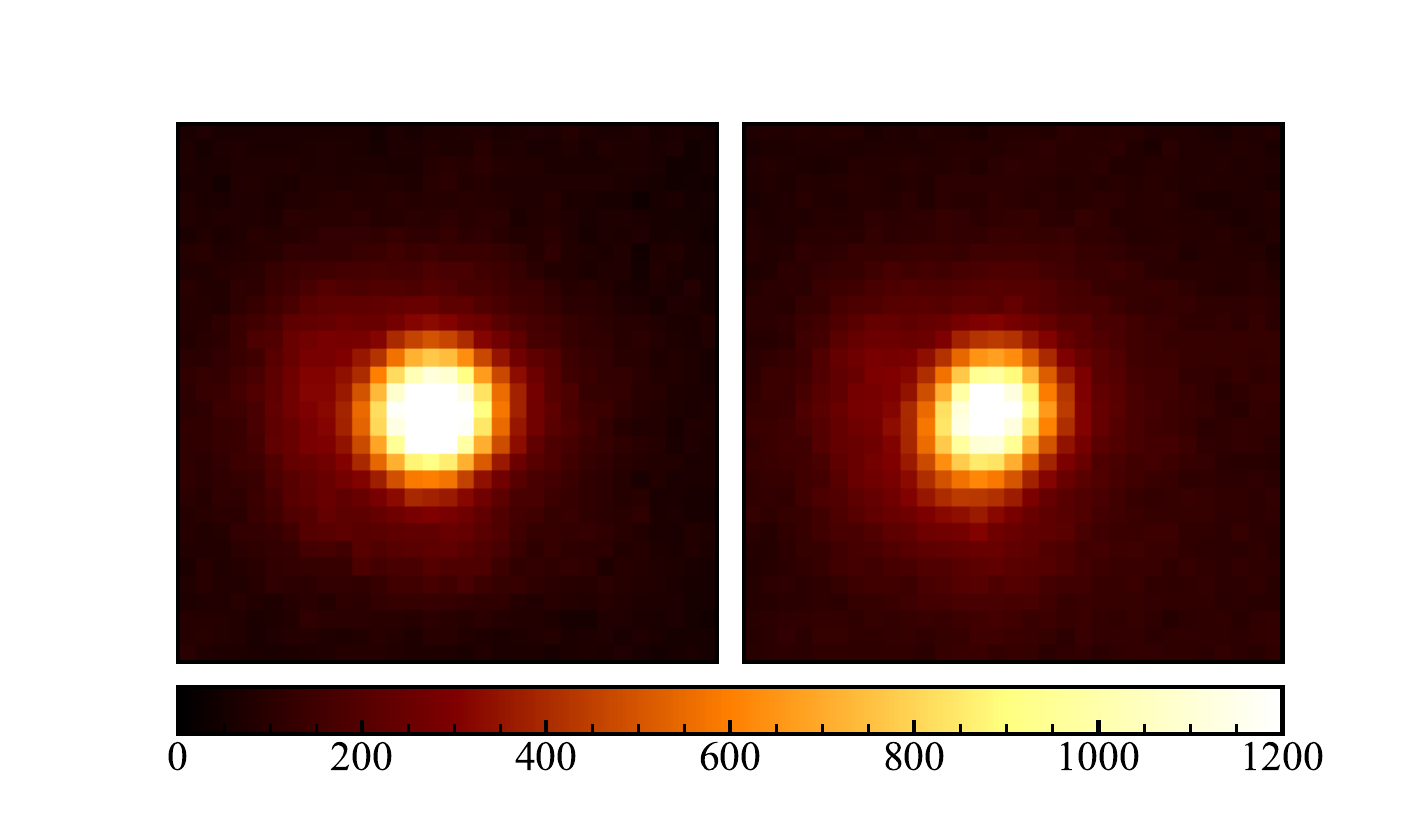}
    \caption{
        Comparison between a representative single star (left) and the target OGLE-2014-BLG-0676 (right) in the same Keck/OSIRIS image. 
        Both panels are displayed using an identical color scale.
    }
    \label{fig:single-vs-target}
\end{figure*}

\rev{
Figure~\ref{fig:single-vs-target} is provided to clarify the structure of the target, which is difficult to discern in Figure~\ref{fig:quad-panel} due to the adopted color scaling. 
By comparing with a nearby single star, we demonstrate that the single star exhibits a symmetric point-spread function, whereas the target shows an elongation.
}

\renewcommand{\arraystretch}{1.4}


\end{document}